\documentclass[lettersize,journal]{IEEEtran}
\usepackage{amsmath,amsfonts}
\usepackage{algorithmic}
\usepackage{algorithm}
\usepackage{array}
\usepackage[caption=false,font=normalsize,labelfont=sf,textfont=sf]{subfig}
\usepackage{textcomp}
\usepackage{stfloats}
\usepackage{url}
\usepackage{verbatim}
\usepackage{graphicx}
\usepackage{cite}
\usepackage{caption}
\usepackage{hyperref}

\usepackage{booktabs}
\usepackage{tabularx}
\usepackage{array}
\usepackage{ragged2e}
\usepackage{makecell}
\usepackage{threeparttable}
\usepackage{caption}
\usepackage{adjustbox}
\usepackage{multirow}
\usepackage{siunitx}

\usepackage{csquotes}

\newcolumntype{L}{>{\RaggedRight\arraybackslash}X}

\begin{document}

\title{Efficient Cross-View Localization in \\6G Space-Air-Ground Integrated Network}

\author{Min Hao, Yanbing Xu, Maoqiang Wu, Jinglin Huang, Chen Shang, \\Jiacheng Wang, Jiawen Kang, Zhu Han, \IEEEmembership{Fellow, IEEE} and Wei Ni, \IEEEmembership{Fellow, IEEE}%

       \IEEEcompsocitemizethanks{
\IEEEcompsocthanksitem Min Hao, Yanbing Xu and Maoqiang Wu are with School of Electronic Science and Engineering, South China Normal University, Foshan, China (e-mail: haomin@scnu.edu.cn, 2024025402@m.scnu.edu.cn, maoqiangwu@m.scnu.edu.cn).
\IEEEcompsocthanksitem Jinglin Huang is with School of Automation, Guangdong University of Technology, Guangzhou, China. He is also with Peng Cheng Laboratory, Shenzhen, China (e-mail: 
2112304047@mail2.gdut.edu.cn).
\IEEEcompsocthanksitem Chen Shang is with School of Electrical and Data Engineering, University of Technology Sydney, Sydney, Australia (e-mail: 
chen.shang@student.uts.edu.au).
\IEEEcompsocthanksitem Jiacheng Wang is with College of Computing and Data Science, Nanyang Technological University, Singapore, Singapore (e-mail: 
jiacheng.wang@ntu.edu.sg).
\IEEEcompsocthanksitem Jiawen Kang is with School of Automation, Guangdong University of Technology, Guangzhou, China (e-mail: 
kjwx886@163.com).
\IEEEcompsocthanksitem Zhu Han is with Electrical and Computer Engineering Department, University of Houston, Houston, United States (e-mail: hanzhu22@gmail.com).
\IEEEcompsocthanksitem Wei Ni is with School of Computing Science and Engineering, University of New South Wales, Sydney, Australia (e-mail: 
wei.ni@ieee.org).

}
}



\maketitle

\begin{abstract}
Cross-view localization (CVL) is increasingly recognized as an effective complement to conventional localization techniques due to its superior coverage and robustness. However, existing CVL methods are largely designed for centralized or loosely distributed processing, making them difficult to scale efficiently in terms of latency, energy consumption, and privacy preservation. At the same time, 6G space-air-ground integrated networks (SAGIN) are expected to provide a globally connected and deeply distributed architecture for future intelligent services. This article explores 6G SAGIN-enabled CVL and highlights a fundamental shift from conventional CVL deployment to network-native distributed intelligence. Specifically, rather than treating SAGIN merely as a communication substrate, we show that its heterogeneous communication and computing resources can fundamentally reshape how CVL is implemented and optimized. We first review the capabilities of CVL and SAGIN, and identify their integration opportunities and representative applications. We then propose a SAGIN-oriented split-inference framework for CVL, which goes beyond conventional split inference by explicitly adapting to the hierarchical and distributed characteristics of space-air-ground networks. Furthermore, we investigate the joint optimization of communication, computation, and confidentiality to establish an efficient and privacy-aware CVL paradigm. Experimental results validate the effectiveness of the proposed framework, and future research directions for 6G SAGIN-enabled CVL are discussed\footnote{\url{https://github.com/xuyanbing11/cross_view_file/tree/main}}. 
\end{abstract}


\section{Introduction}
Vision-based localization serves as an important complement to high-precision positioning in 6G networks. In fields such as autonomous vehicles, unmanned aerial vehicles (UAVs), and augmented reality (AR), vision-based localization utilizes image or video information to determine the position and orientation of targets and has been widely adopted\cite{11142299}. In particular, in scenarios with weak satellite signals, such as urban canyons, indoor environments, and low-altitude UAV operations, vision-based localization offers significant advantages over traditional global navigation satellite system (GNSS)- or inertial measurement unit (IMU)-based methods. Despite significant progress in vision-based localization, existing methods largely rely on images captured from similar viewpoints and scales, which is insufficient to meet the intelligent, immersive, and high-precision localization requirements envisioned for 6G applications.


To overcome the limitations of same-view localization, cross-view localization (CVL) has emerged as a promising paradigm\cite{10644040}. CVL performs feature matching between a given query image and a database of reference images captured from different viewpoints or altitudes. It thereby enables the estimation of the absolute position and orientation of the query image, serving as a complement to high-precision localization rather than a replacement for dedicated fine-grained positioning modules. A comparison of the advantages and limitations between CVL and same-view localization is presented in Table \ref{tab:vision_fusion_clean}. CVL is capable of supporting multi-scale and multi-view localization, and enhances robustness to diverse scenarios. However, large cross-domain heterogeneity remains a major challenge for CVL. Significant differences in viewpoint, resolution, and imaging modalities across space, air, and ground visual data make feature matching highly challenging. Moreover, CVL suffers from privacy leakage, insufficient computational power, and high latency when processing multi-source data\cite{11602083}.

\begin{table*}[t!]
\centering
\caption{Comparison of Visual Data Sources and Cross-Domain Fusion for Localization}
\label{tab:vision_fusion_clean}
\begin{threeparttable}
\small              
\renewcommand{\arraystretch}{1.18}
\setlength{\tabcolsep}{4pt}
\begin{tabularx}{\textwidth}{lLLLL}
\toprule
\textbf{Dimension} & \thead{Satellite\\Images} & \thead{UAV\\Images} & \thead{Ground\\Images} & \thead{Tri\textendash fusion\\(Cross-View Loc.)} \\
\midrule
Viewpoint & Nadir (vertical) & Top-down + oblique & Horizontal / first-person & Multi-view fusion \\
Resolution & Meter-level & Centimeter-level & Sub-meter or finer & Cross-scale integration \\
Dynamics & Static & Highly dynamic & Moderately dynamic & Multi-dynamic adaptation \\
Visual Coverage\tnote{a} & 10 km & $ \le $1 km & Local area & Global-to-local coverage \\
Advantages & Wide-area positioning & High resolution; flexible & Rich fine-grained details & High precision; robustness \\
Limitations & Refresh slowly & Unstable features & Severe occlusion & Large cross-domain heterogeneity \\
Applications & Digital twins & UAV inspection; low-altitude economy & V2X / connected vehicles; AR nav. & Autonomous driving; 6G \\
\bottomrule
\end{tabularx}

\begin{tablenotes}
\footnotesize
\item[a] The "Coverage" values refer to the representative ground scene coverage of a single image in the CVL context, rather than communication coverage or orbital coverage. They are used only for qualitative comparison among different image sources.
\end{tablenotes}

\end{threeparttable}
\end{table*}

Motivated by the potential benefits of 6G, this article investigates CVL in the context of 6G space-air-ground integrated networks (SAGIN). The SAGIN is constructed by integrating satellite communications, high-altitude platforms, and terrestrial networks, thereby enabling global coverage, seamless connectivity, and low-latency ubiquitous communication infrastructure\cite{11053513}. Using CVL in 6G SAGIN brings several notable advantages. Although GNSS signals may be unreliable in complex environments, fusing multi-view information significantly improves localization reliability and robustness, thus making the system more resilient to occlusion and environmental changes. Moreover, combining global references from aerial and satellite imagery with local details from ground-level images can enhance the  positioning accuracy. In addition, CVL can perform forward feature extraction at edge nodes within 6G SAGIN, requiring only the transmission of low-dimensional features and thus reducing communication overhead. Finally, 6G SAGIN inherently possess distributed computing and communication capabilities, making them inherently compatible with the hierarchical processing of CVL\cite{10260323}.

To address the fundamental challenges of deploying CVL in large-scale, dynamic, and privacy-sensitive environments, we investigate CVL from the perspective of 6G SAGIN. In this work, CVL mainly localizes UAVs and vehicles in SAGIN scenarios. For UAV localization, vehicle-view and satellite-view images can serve as references. For vehicle localization, UAV-view and satellite-view images can serve as references. Conventional CVL methods rely on centralized or loosely distributed processing, while existing split-inference solutions are not tailored to cross-domain visual localization over hierarchical networks. To address these limitations, we develop a network-native distributed framework for SAGIN-enabled CVL. Specifically, the main contributions are summarized as follows:
\begin{itemize}
    \item We present an overall architecture for 6G SAGIN-enabled CVL and highlight its fundamental integration logic. In particular, we show that SAGIN can reshape CVL from a stand-alone visual task into a distributed and communication-aware intelligent service. Based on this architecture, we identify several key research directions, including visual processing, privacy preservation, and communication-computation coordination.

    \item We propose a space-air-ground collaborative split-inference framework for CVL. The framework supports cross-domain feature extraction and matching by leveraging distributed communication and computing resources across heterogeneous network segments. Moreover, it goes beyond conventional split inference by explicitly considering the tightly coupled trade-off among latency, energy consumption, and privacy preservation.

    \item We validate the proposed framework through a case study under different image quantities and privacy attack settings. Experimental results show that the framework can improve localization accuracy and data security, while confirming the necessity of jointly optimizing communication, computation, and confidentiality for efficient SAGIN-enabled CVL.
\end{itemize}

The rest of the article is organized as follows: An overview of CVL and 6G SAGINs is presented in Section \ref{II}. Section \ref{III} describes the proposed split-inference framework and optimization scheme, while Section \ref{IV} presents the experimental evaluation of the proposed methods. We discuss future research directions in Section \ref{V}. Finally, we conclude our article in Section \ref{VI}.

\section{Overview of Cross-View Localization and SAGIN}\label{II}

\subsection{Cross-View Localization}
\begin{figure*}[!t]
    \centering
    \includegraphics[width=1\textwidth]{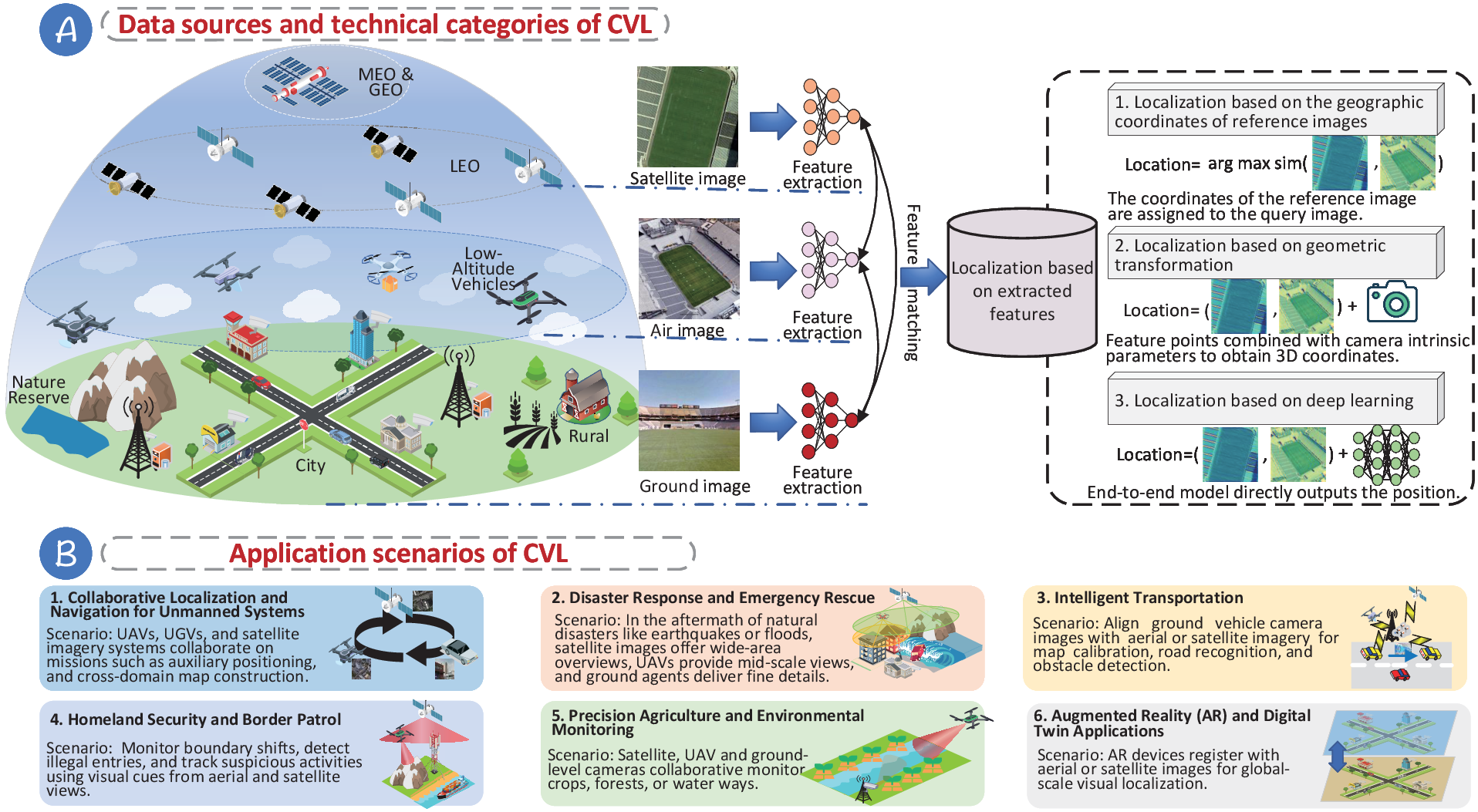} 
    \caption{Overview of Cross-View Localization. 
    illustrates the data sources and localization methods of CVL. 6G SAGIN provides multi-source image data for CVL, which enables localization through different methods. \textit{Part B} presents the application scenarios of CVL from different perspectives.}  
    \label{fig1}         
\end{figure*}

CVL is an emerging technique that estimates the absolute position and orientation of a query image by matching it with reference images captured from different viewpoints, altitudes, or modalities\cite{11141471}. As shown in Fig.~\ref{fig1}, CVL leverages heterogeneous visual data sources within space-air-ground domains to achieve localization, particularly in scenarios where accurate GNSS signals cannot be received due to building obstructions or natural disasters. CVL typically involves three main stages:
\begin{itemize}
\item \textbf{Feature Extraction:} Local or global features are extracted from both query and reference images using traditional descriptors or deep learning-based models.
\item \textbf{Feature Matching:} Extracted features are aligned across multi-view or cross-modal datasets to establish visual correspondences, often enhanced through semantic cues or multi-modal fusion.
\item \textbf{Localization:} The absolute position can be determined by either assigning the geographic coordinates of the best-matching reference image as an approximate localization result, or performing geometric transformations for relative pose estimation. Alternatively, end-to-end localization models can be directly designed and trained, where the input is an image and the output is the estimated location.
\end{itemize}

Feature extraction is a critical step in CVL, and different studies define features in different ways, which can be broadly categorized as follows:
\begin{itemize}
\item \textbf{Image-feature-based methods:} The core idea is to extract local or global features from images and identify corresponding points across cross-domain data. Representative methods include scale-invariant feature transform (SIFT), oriented fast and rotated brief (ORB), and network vector of locally aggregated descriptors (NetVLAD)\cite{9939020}. These methods are theoretically well-established and provide strong algorithmic interpretability. However, their localization accuracy is low when dealing with large viewpoint variations or substantial differences in resolution.
\item \textbf{Semantic-feature-based methods:} With the advancement of artificial intelligence, the semantic information of images has attracted increasing attention. The core idea is to leverage deep learning to extract cross-view-invariant semantic features, enabling matching and localization. Adaptability to cross-view and cross-scale variations is one of their main advantages. However, the models heavily depend on large volumes of labeled data.  Representative methods include convolutional neural network (CNN), residual network (ResNet), and Vision Transformer\cite{10852334}.
\item \textbf{Multi-modal feature fusion-based methods:} CVL can also be realized by incorporating information from other modalities. The core mechanism is to integrate multi-source data such as visual, light detection and ranging (LiDAR), IMU, and GNSS for localization. These methods enable the joint extraction of semantic and geometric features, offering high robustness. However, the overall system complexity increases, necessitating latency optimization. Existing methods include contrastive language–image pre-training (CLIP), local feature transformer (LoFTR), and self-distillation with no labels v2 (DINOv2)\cite{11105551}.
\end{itemize}

From the characteristics of the aforementioned methods, the core features of CVL can be identified as cross-domain fusion, high-accuracy robustness, and native adaptability to 6G. 
In the scenarios where GNSS signals are weak or unavailable, CVL can still perform reliable localization and leverage multi-view data to compensate for the degradation in localization accuracy due to low image resolution. Meanwhile, CVL is inherently well-suited to high-bandwidth, low-latency, and multi-modal big data scenarios. Based on the above characteristics, Fig.~\ref{fig1} illustrates the potential application scenarios of CVL.

\subsection{6G Space-Air-Ground Integrated Networks}
SAGIN unifies high-altitude platforms (such as satellites), low-altitude aerial platforms (such as UAVs or balloons), and ground platforms (such as cellular networks or vehicles) into a single 6G network. In this paper, UAVs denote aerial platforms rather than aerial base stations. The 6G SAGIN aims to achieve global coverage, ubiquitous connectivity, ultra-reliability with low latency, integrated sensing and communication, and sustainability\cite{10409745,10745905}.

There are numerous application scenarios for 6G SAGIN, and an example is illustrated using a UAV-based urban delivery system. In the low-altitude economy, UAVs are expected to undertake tasks such as parcel delivery, emergency supply transportation, and urban infrastructure inspection. By contrast, UAVs operating in urban canyon environments face challenges such as GNSS signal outages and communication interruptions\cite{10770154}. 6G SAGIN addresses these challenges by integrating satellites (Space), low-altitude platforms (Air), and terrestrial networks (Ground) as follows: 
\begin{itemize}
\item Before flight, UAVs interact with the air traffic management system to download 3D flight routes based on the urban digital twin. A SAGIN allocates multi-domain resources to ensure prioritized availability of communication links. 
\item During flight, the UAVs continuously receive signals from low earth orbit (LEO) satellites, low-altitude relay platforms, and ground roadside units (RSUs). When entering densely built-up areas, CVL is activated, allowing the UAVs to share point cloud and visual data with the RSUs to achieve sub-meter-level obstacle avoidance.  
\item After completing the flight, the UAVs upload their flight paths, communication logs, and energy consumption data to the urban digital twin, enabling subsequent simulation and optimization. 
\end{itemize}

\begin{figure*}[!t]
    \centering
    \includegraphics[width=1\textwidth]{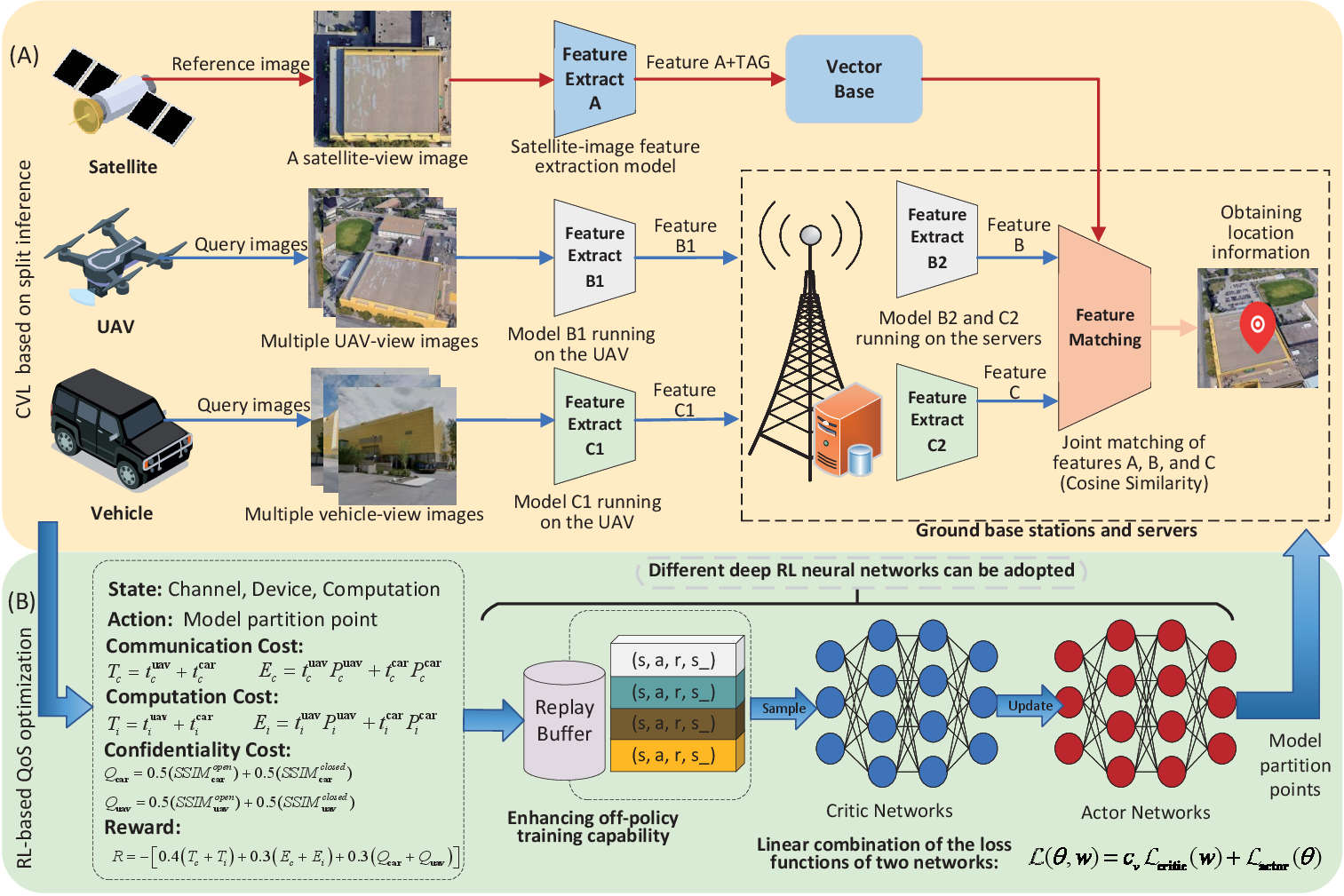} 
    \caption{Split-inference framework for CVL in 6G SAGIN. \textit{Part A} illustrates the process of implementing CVL through split-inference within the 6G SAGIN. \textit{Part B} presents the Tri-Co optimization scheme based on reinforcement learning (RL). In the considered architecture, UAVs or vehicles initiate the localization procedure by submitting query images. Satellite-view images mainly provide feature and geographic tag references, while the final feature matching and location estimation are performed by servers connected to ground base stations.}
    \label{fig2}         
\end{figure*}
\section{SAGIN-Enabled  Cross-View Localization}\label{III}

\subsection{Requirement}
Achieving high-quality CVL requires robust communication infrastructure, and CVL should be natively embedded within 6G networks so that its implementation is tightly integrated with the network architecture. Within the communication and computation architecture of 6G networks, CVL can be optimized from the following aspects:
\begin{itemize}

\item \textbf{Architecture Alignment:} 6G emphasizes a fully integrated, ubiquitous, and intelligent network architecture spanning space, air, ground, and sea. The research direction of CVL should align with this development trend by pursuing higher localization accuracy and adopting more reliable methods to seamlessly integrate into the 6G network ecosystem.

\item \textbf{Distributed Inference:} To improve efficiency, 6G will be developed based on a fully distributed infrastructure that integrates communication, computation, sensing, and storage, and the implementation of CVL will likewise be distributed. Splitting feature-matching models for distributed training and inference is a key approach to enhancing the efficiency of CVL.

\item \textbf{Quality of service optimization:} As an important task within 6G networks, CVL requires further optimization of its quality of service (QoS). In particular, when using split-inference for feature matching, the communication latency, computational energy consumption, and data confidentiality cost across space-air-ground devices need to be jointly optimized.

\end{itemize}

\subsection{Proposed Framework}

\subsubsection{CVL based on Split-Inference}Split-inference is an AI model deployment technique in which a neural network is divided into two (or more) parts and executed across different devices. Typically, the front part of the model runs on a resource-constrained edge device to extract intermediate features, while the remaining part runs on a more powerful edge server or cloud server. Different from ordinary visual tasks, split point selection in CVL should also preserve cross-view feature matching capability.
Part (a) of Fig.~\ref{fig2} illustrates the CVL process based on split-inference.

The image data in Fig.~\ref{fig2} are sourced from satellites, UAVs, and vehicles across the space-air-ground domains. Satellite images serve as reference images, from which features are extracted and associated with corresponding geographic coordinate tags. The features of satellite images and their associated tags are pre-stored in vector form on a ground server, and organized into a database to facilitate on-demand retrieval for CVL tasks. Images captured by UAVs and vehicles can both serve as query images for CVL. Alternatively, any of them can be used as a reference image while the rest functions as a query image. The acquisition cost of satellite imagery is relatively high, whereas the imaging cost of air and ground devices is much lower. Therefore, CVL can achieve localization by selecting a single satellite image together with multiple images captured within the Earth's lower atmosphere and surface.

The feature extraction models for satellite-view, UAV-view, and vehicle-view images are implemented as three neural networks with identical architectures but different parameters (e.g., CNN or Transformer). During model training, each location is treated as a distinct class, and the outputs of the three neural networks are then fed into a shared linear classifier. This constrains images from different viewpoints to the same discriminative space, thereby achieving cross-view alignment. During model inference, the feature extraction model for satellite images is executed on servers aboard LEO satellites. The extracted satellite features, together with their geographic tags, are then transmitted to the ground-side server/database for subsequent retrieval and matching, instead of sending raw images directly. The feature extraction models for UAVs and vehicles are divided into two parts: the front part is executed on the terminal devices (UAVs and vehicles), while the back part runs on a ground base station. This design conserves the computational resources of UAVs and vehicles, thereby reducing both computational and communication energy consumption. Moreover, since images captured by UAVs and vehicles may involve sensitive information such as faces, license plates, and restricted areas, transmitting intermediate feature data can effectively mitigate privacy leakage risks. 

After feature extraction, the features from the three types of images are jointly matched. The servers connected to ground base stations obtain features from images of different viewpoints and perform nearest-neighbour retrieval using cosine similarity, where a similarity greater than a predefined threshold is regarded as a successful match. Once feature matching is successfully achieved, the query images can obtain an approximate location by associating them with the geographic coordinates of the corresponding reference images, thereby providing an auxiliary positioning cue for subsequent fine localization. The above framework is broadly applicable to CVL tasks within 6G SAGIN, while also accounting for limited device computing capabilities and ensuring the protection of sensitive data.

\begin{figure*}[!t]
    \centering
    \includegraphics[width=1\textwidth]{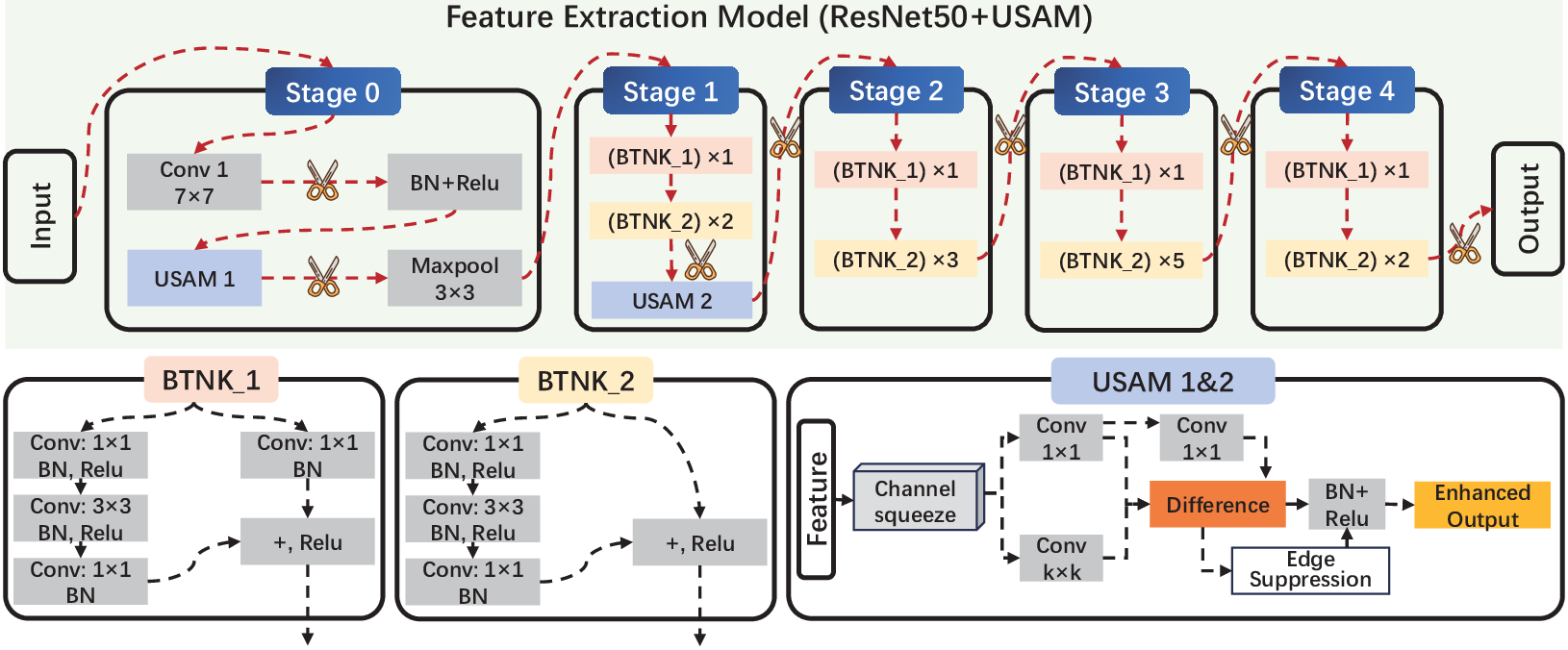} 
    \caption{The internal structure of the feature extraction model. The scissor mark indicates the model partition point defined in this work. The USAM compresses channel features into a spatial map, extracts multi-branch spatial information, and generates an attention map via difference fusion.}
    \label{f_e}         
\end{figure*}

\subsubsection{Tri-Co-Optimization Scheme}
In the above framework, to achieve a better user experience, joint optimization across communication, computation, and confidentiality is required.
\begin{itemize}

\item \textbf{Communication:} The communication links for collecting multi-source images required by CVL are long distance and highly heterogeneous. Various types of image data must be transmitted among satellites, UAVs, ground devices, and a cloud server, where even slight data redundancy can result in network congestion. In split-inference, the selection of model partition points determines the size of intermediate features, which in turn affects the latency and energy consumption during transmission. Additionally, for UAVs and vehicles issuing localization requests through CVL, optimizing communication energy consumption is crucial for reducing overall energy usage. 

\item \textbf{Computation:} In CVL, performing feature extraction and feature matching on high-resolution images incurs substantial computational resource consumption. However, the computing capabilities of UAVs and vehicles are limited, and excessive computational loads can lead to rapid battery depletion and increased device overheating. Moreover, feature extraction in CVL increasingly involves multi-modal feature fusion, which further increases computational complexity. Although split-inference can alleviate the computational burden on terminal devices, optimization of computing strategies is still required to support concurrent multitasking on these devices.

\item \textbf{Confidentiality:} In the 6G era, CVL systems will collect and transmit terabytes of visual data, and directly uploading raw images would pose significant risks to user privacy, commercial confidentiality, and national security. If all raw images are transmitted to the cloud, large-scale privacy data could be compromised through potential cloud-based attacks. The split-inference framework can avoid directly uploading raw data, yet original data can still be partially reconstructed through open-box or closed-box attacks. Therefore, optimizing the data confidentiality of CVL is also essential.

\end{itemize}

The above analysis highlights the necessity of optimizing Tri-Co. However, optimizing a single dimension in isolation is insufficient and a joint optimization of Tri-Co is required. In particular, within the CVL architecture based on split-inference, the selection of partition points directly affects Tri-Co. When the model partition point is placed deeper, the UAV or vehicle extracts higher-dimensional image features, and the size of the intermediate features tends to decrease. This can reduce the communication energy consumption and latency of terminal devices, and also enhance the confidentiality of raw data, but the computational energy consumption increases. When the model partition point is placed earlier, the UAV or vehicle extracts lower-dimensional image features, and the size of the intermediate features tends to increase. This results in increased communication energy consumption and latency on terminal devices, reduced confidentiality of raw data, and lower computational energy consumption.

Part (b) of Fig.~\ref{fig2} illustrates the QoS optimization scheme based on RL, which performs the joint optimization of Tri-Co. First, the state of the 6G SAGIN is recorded, including channel parameters, device battery status, computing resource availability, and other environmental information relevant to CVL. Subsequently, the configurations of the feature extraction models in CVL are recorded, including the model partition points used by UAVs and vehicles. Since UAVs and vehicles differ in computing capabilities, battery capacities, and other parameters, the selected model partition points may vary accordingly. The overall optimization objective of the system is to minimize communication, computation, and confidentiality costs. Accordingly, the system’s reward function can be defined as the weighted sum of these three costs, and maximizing the additive inverse of this sum achieves the maximization of the system reward. Fig.~\ref{fig2} illustrates an off-policy deep RL network, where experience replay is employed to maximize sample utilization and improve training stability. The Critic network outputs evaluations of actions, while the Actor network generates model partitioning strategies and applies them to CVL. The joint optimization of Tri-Co is not limited to off-policy RL methods. We can adopt policy-based RL learning or other optimization approaches depending on practical requirements.

\section{Case Study}\label{IV}


 \subsection{Experimental Configurations} 
In the case study, three experiments are designed as follows:
\begin{itemize}

\item \textbf{Experiment 1:} The feasibility and localization performance of UAVs or vehicles using CVL based on 6G SAGIN are evaluated through simulation experiments. The experiment is configured with varying numbers and viewpoints of images for matching.

\item \textbf{Experiment 2:} By simulating open-box and closed-box attacks, the effectiveness of protecting raw data at different model partition points in split-inference is evaluated. The experiment conducts a quantitative evaluation of the reconstructed images.

\item \textbf{Experiment 3:} A joint optimization problem for communication, computation, and confidentiality in the QoS metrics of CVL is formulated. The formulated optimization problem is solved using a range of RL algorithms, and the convergence performance of different algorithms is compared.

\end{itemize}

Specifically, in Experiment 1, we adopt ResNet-50 as the backbone feature extraction model, with two unified spatial attention modules (USAM) incorporated to enhance feature matching performance. The internal structure of the feature extraction model is shown in Fig.~\ref{f_e}. The USAM incorporates uncertainty information to dynamically adjust the attention distribution, thereby enhancing the robustness of feature selection. We insert USAM modules at Stage 0 and Stage 1 of ResNet-50, respectively. The simulation scenarios cover localization for both UAVs and vehicles. For UAV localization, we use satellite-view and vehicle-view images as reference images, whereas for vehicle localization, satellite-view and UAV-view images serve as the references. The dataset used is University-1652, which contains satellite, UAV, and vehicle-view images of 1,652 buildings from 72 universities\cite{10.1145/3394171.3413896}. In the experiments of this work, one satellite image, four UAV images, and four vehicle-view images were selected for each building, and the impact of varying image quantities on localization accuracy was investigated. We employ Recall@K and average precision (AP) to evaluate the effectiveness of feature matching, thereby assessing the performance of CVL. We use Recall@K to measure whether the system successfully retrieves the correct matching image within the top-K results. If the true matching building image appears among the top K retrieved results, the Recall@K is set to 1; otherwise, it is 0. Since Recall@K cannot handle multiple matches, we also adopt AP in the experiments to provide a more comprehensive evaluation of localization performance. The AP is calculated by averaging the precision values at the ranks where each true matching image appears. A higher AP indicates that more true matches are retrieved earlier, reflecting better localization performance.



\newcommand{\firstcolw}{2.3cm}        
\newcommand{\gapcolw}{0.00cm}          
\newcommand{\panelfont}{\footnotesize} 


\begin{table*}[t!]
\centering
\renewcommand{\arraystretch}{1.18}
\setlength{\tabcolsep}{3pt}

\begin{threeparttable}
\caption{CVL Matching in a Representative Air-Ground Multi-View Scenario on the University-1652 Dataset, based on a Single Satellite-view Reference Image (\%).}
\label{uav_ground}

\newcommand{\tablesubtitle}[1]{%
  \addlinespace[0.35em]
  \multicolumn{11}{c}{{\panelfont\textbf{#1}}}\\[-0.25em]
  \cmidrule(lr){1-11}
}

\newcommand{\metrichead}{%
\multicolumn{1}{c}{Recall@1} & \multicolumn{1}{c}{Recall@5} &
\multicolumn{1}{c}{Recall@10} & \multicolumn{1}{c}{Recall@top1} &
\multicolumn{1}{c}{AP}
}

\begin{tabular*}{\textwidth}{@{\extracolsep{\fill}}@{}%
  p{\firstcolw} @{\hspace{\gapcolw}} *{10}{S} @{}}
\tablesubtitle{(a) \, Limited number of vehicle-view images
}
\multirow{2}{\firstcolw}[-0.9ex]{\centering\textbf{Air$\leftrightarrow$Ground}} &
\multicolumn{5}{c}{1 vehicle-view image
} & \multicolumn{5}{c}{2 vehicle-view images} \\
\cmidrule(lr){2-6}\cmidrule(lr){7-11}
& \metrichead & \metrichead \\
\midrule
1 UAV-view image & 47.08 & 72.04 & 80.88 & 81.74 & 52.91 & 51.21 & 75.89 & 83.74 & 84.02 & 56.76 \\
2 UAV-view images & 48.36 & 72.75 & 81.31 & 82.03 & 54.02 & 51.36 & 76.89 & 84.59 & 85.16 & 57.06 \\
3 UAV-view images & 47.65 & 73.18 & 81.60 & 82.88 & 53.40 & 51.64 & 77.89 & 84.45 & 85.59 & 57.49 \\
4 UAV-view images & 48.36 & 73.89 & 82.45 & 83.59 & 54.09 & 52.21 & 78.46 & 85.45 & \textbf{85.88} & \textbf{58.11} \\
\bottomrule
\end{tabular*}

\par\addvspace{0.75\baselineskip}

\begin{tabular*}{\textwidth}{@{\extracolsep{\fill}}@{}%
  p{\firstcolw} @{\hspace{\gapcolw}} *{10}{S} @{}}
\tablesubtitle{(b) \, Increasing the number of vehicle-view images}
\multirow{2}{\firstcolw}[-0.9ex]{\centering\textbf{Air$\leftrightarrow$Ground}} &
\multicolumn{5}{c}{3 vehicle-view images} & \multicolumn{5}{c}{4 vehicle-view images} \\
\cmidrule(lr){2-6}\cmidrule(lr){7-11}
& \metrichead & \metrichead \\
\midrule
1 UAV-view image & 53.64 & 77.46 & 83.74 & 84.17 & 59.07 & 54.21 & 78.03 & 84.59 & 84.88 & 59.61 \\
2 UAV-view images & 53.64 & 77.89 & 85.02 & 86.16 & 59.21 & 55.06 & 78.74 & 85.16 & 86.45 & 60.40 \\
3 UAV-view images & 54.78 & 78.82 & 86.86 & 87.02 & 60.17 & 56.09 & 79.03 & 85.88 & 86.45 & 61.26 \\
4 UAV-view images & 55.63 & 79.74 & 85.88 & 86.88 & 60.95 & 55.92 & 79.89 & 86.31 & \textbf{86.88} & \textbf{61.35} \\
\bottomrule
\end{tabular*}

\end{threeparttable}
\end{table*}


\begin{figure*}[!t]
  \centering
  \subfloat[Attack effects of open-box and closed-box approaches\label{wb}]{
    \includegraphics[width=0.45\linewidth]{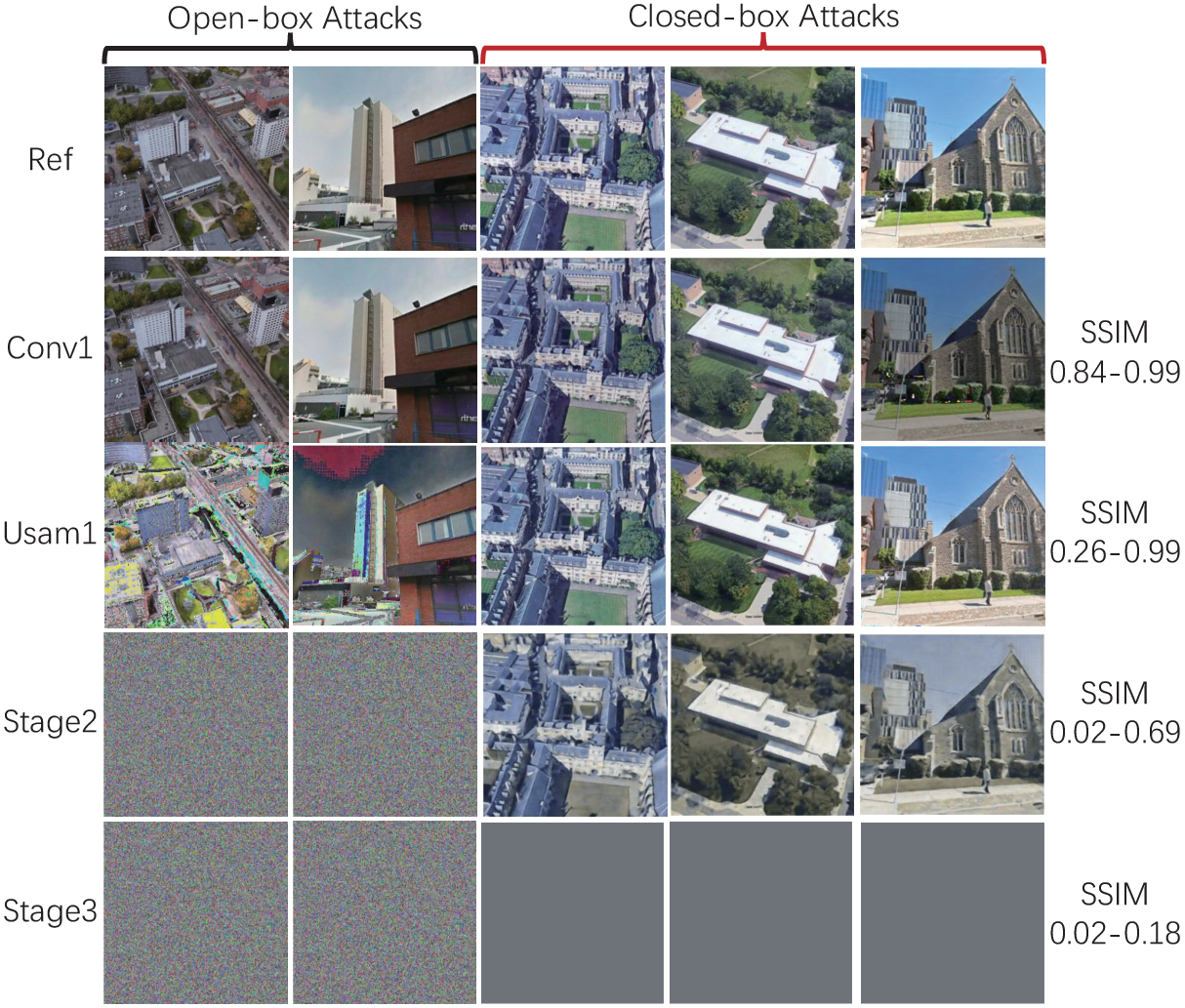}%
  }\hfil
  \subfloat[Comparison of the convergence performance of different RL algorithms\label{rl}]{
    \includegraphics[width=0.45\linewidth]{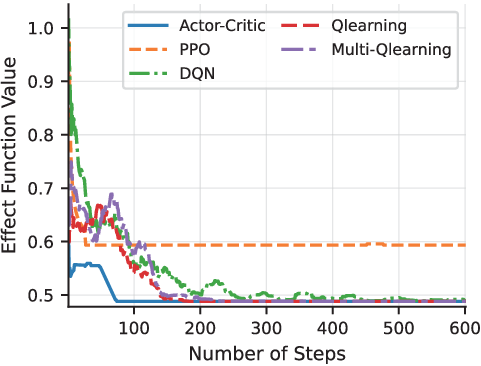}%
  }
  \caption{Data security of CVL with split-inference and the joint optimization performance of Tri-Co.}
  \label{wb+rl}
\end{figure*}

In Experiment 2, open-box and closed-box attacks are carried out on models with different partition points\cite{10005100}. The open-box attack refers to a scenario in which the attacker has full knowledge of the target model, including its architecture, parameter weights, and gradient information, and exploits this knowledge to infer the original data, thereby causing privacy leakage. The closed-box attack refers to a scenario in which the attacker only has access to intermediate features and the distribution of the original data, and attempts to steal privacy information by training an inversion model to reconstruct the original data. In Fig.~\ref{f_e}, the selected partition point is indicated by a scissor icon. We apply this partitioning strategy identically to both UAVs and vehicles. To evaluate the difference between the reconstructed images from the two types of attacks and the original images, we use the structural similarity index (SSIM) for quantitative evaluation. The SSIM focuses on structural information by jointly considering luminance, contrast, and structural similarity. Its value ranges from 0 to 1, with values closer to 1 indicating higher similarity between two images, and thus greater exposure of privacy information.

In Experiment 3, we use the NVIDIA Quadro P400 to simulate the computing platform of a UAV. It provides a peak FP32 performance of 0.641 TFLOPS with a maximum power consumption of 30W. Its single-slot design and ultra-low power consumption make it well-suited for lightweight professional graphics and computing tasks. We then use the NVIDIA DRIVE AGX Xavier to simulate the computing platform of a vehicle. Its GPU delivers 1.3 TFLOPS of single-precision (FP32) performance. The wireless communication cost between UAVs or vehicles and ground servers primarily considers latency and energy consumption. The selection of model partition points directly affects the size of the intermediate features to be transmitted. By performing a forward pass at the split point and examining the shape of the output tensor at that layer, the dimensions and size of the intermediate features can be determined, thereby enabling the calculation of the latency and energy consumption required for wireless communication with a ground server. The computation cost primarily considers latency and energy consumption of UAVs and vehicles during split-inference. The selection of model partition points leads to different computational workloads on terminal devices (e.g., UAVs or vehicles), which in turn results in variations in latency and energy consumption. Based on the power consumption of floating-point operations for different devices, the corresponding computational latency and energy consumption can be obtained. The cost of confidentiality is primarily evaluated based on SSIM. The selection of partition points in the feature extraction model directly influences the quality of images reconstructed by open-box and closed-box attack models, thereby affecting the confidentiality of the original data. The confidentiality cost of the UAV and vehicle is defined as the weighted sum of the SSIM values under two attack models, where the weights are set to 0.5.

After obtaining the three cost functions of communication, computation, and confidentiality for the CVL system, we formulate the optimization problem. The partition point selection is treated as an optimization variable, and the CVL system aims to minimize the weighted sum of latency, energy consumption, and confidentiality cost. The partition point affects not only transmission and computation costs, but also the quality of cross-view features for matching. As shown in Fig.~\ref{fig2}, the weights of latency, energy consumption, and confidentiality are set to 0.4, 0.3, and 0.3, respectively. We assign a higher weight to communication to prioritize latency in meeting the timeliness requirements of the CVL task, while energy consumption and data security are equally considered. Several fundamental RL algorithms are applied to optimize the above problem, providing references for future research. The main algorithms include Actor-Critic, Q-Learning, Multi-Q-Learning, DQN, and PPO, all of which can serve as baselines for comparison with other algorithms\cite{pmlr-v48-mniha16}.

\subsection{Performance Analysis and Scalability Discussion} Table \ref{uav_ground} presents the matching performance between UAV-view and vehicle-view images under different settings. The results indicate that performance improves consistently as the number of UAV-view and vehicle-view images increases. In the limited vehicle-view scenario, adding more UAV-view images enhances recall and AP, with the best performance observed when four UAV-view images are combined with two vehicle-view images. When the number of vehicle-view images is further increased, the improvement becomes more significant, and the highest accuracy is achieved with four UAV-view and four vehicle-view images (Recall@top1 = 86.88\%, AP = 61.35\%). These findings highlight the critical role of multi-view integration in enhancing cross-view matching accuracy and robustness. These findings also demonstrate that CVL based on 6G SAGIN offers broad opportunities for further research. For example, future work should examine how to select from the massive volume of images provided by 6G SAGIN to improve CVL accuracy, and how to integrate multi-view images to enhance CVL efficiency.

Fig.~\ref{wb} demonstrates the attack effects of open-box and closed-box approaches at different feature extraction stages. The first row presents the original images as a reference, while the subsequent four rows display the reconstruction results of the two attack approaches under different partition points in split-inference. Under open-box attacks, the difficulty of reconstructing the original images increases as the model partition point becomes deeper, resulting in images at Stage 2 and Stage 3 being almost completely corrupted and visually unrecognizable. In contrast, closed-box attacks preserve more structural content, but image fidelity still degrades progressively with increasing depth. The SSIM values on the right-hand side quantitatively confirm this trend: while shallow partition points (Conv1) maintain relatively high similarity scores (0.84–0.99), deeper partition points cause the similarity to drop sharply, reaching as low as 0.02–0.18 at Stage 3. Fig.~\ref{wb} does not present the reconstruction results for the partition point at Stage 4, as the images can no longer be recovered at this stage. These results demonstrate that as the partition point moves deeper into the neural network, the confidentiality of the original data becomes stronger. Naturally, the communication and computation resources also change accordingly.

Fig.~\ref{rl} compares the convergence performance of different RL algorithms in terms of effect function value over training steps. The horizontal axis represents 600 sampled training steps, while the vertical axis denotes the effect function value, defined as the weighted sum of communication, computation, and confidentiality costs. The results show that the Actor-Critic method achieves the fastest and most stable convergence, reaching a lower effect function value of around 0.48 with minimal fluctuations. The Actor-Critic model adopts a classical dual-network architecture. The learning rate of the Actor is set to 0.0003, while the Critic is set to 0.0007. A discount factor of 0.99 is used to quantify the importance of future rewards, and the advantage function is computed based on TD error. Q-Learning also converges but exhibits slower reduction and greater variability in effect function value. In contrast, DQN, PPO, and Multi-Q-learning demonstrate unstable convergence behaviour and remain at higher function values, indicating inferior optimization performance. Since the RL state includes channel, device, and computing conditions, the learned policy adapts to varying system states. Actor-Critic shows faster and more stable convergence. Broader sensitivity and generalization evaluation is left for future work.

The current case study considers a limited number of UAVs and vehicles to verify the feasibility of SAGIN-enabled CVL. When many devices simultaneously request CVL services, the main bottleneck is expected to move to server-side feature retrieval and matching. In the proposed split-inference framework, UAVs and vehicles only extract front-end features and transmit intermediate representations, while final matching and localization are completed at ground servers. This design reduces raw-image transmission and terminal-side computation. However, concurrent requests may still increase matching latency because of server workload accumulation, queueing delay, and wireless resource contention. Large-scale concurrent evaluation should be studied in future work.

\section{Future Directions}\label{V}

\subsection{Multi-Source Visual Fusion and Cross-Modal Feature Alignment} Future research is expected to develop advanced strategies for multi-source visual fusion and cross-modal feature alignment. By harmonizing heterogeneous imagery across space, air, and ground views, feature representations can be unified to improve CVL. Lightweight fusion models and self-supervised adaptation will enhance robustness against modality gaps, enabling scalable and reliable localization systems.

\subsection{SAGIN Based CVL Datasets and Network-Aware Evaluation} Publicly available datasets for space-air-ground CVL remain limited. Existing benchmarks mainly support visual matching and localization evaluation, but cannot fully reflect the network complexity of practical 6G SAGIN systems. A key future direction is to develop network-aware datasets and evaluation platforms that jointly model multi-view images and network conditions, enabling more realistic validation of SAGIN-oriented CVL frameworks.

\subsection{Privacy-Preserving and Secure Trusted Localization} Ensuring privacy-preserving and secure trusted localization is critical for practical deployment. Future research can integrate cryptographic primitives, distributed learning, and verifiable computation frameworks. These approaches will protect sensitive visual data, enhance system accountability, and strengthen resilience against adversarial attacks, while maintaining localization accuracy in diverse mission-critical scenarios.



\section{Conclusion}\label{VI}
In this paper, we have investigated efficient CVL based on SAGIN, focusing on how CVL can be integrated with the 6G network architecture. First, we have provided an overview of CVL and 6G SAGIN. We have conducted a comparison between single-view and cross-view image localization. We have also summarized the fundamental methods and potential applications of CVL. Subsequently, we have proposed a split-inference-based CVL framework within the 6G SAGIN architecture. We have designed this framework to enhance the efficiency and accuracy of cross-domain visual localization for low-altitude and ground devices by integrating the communication and computation resources of space-air-ground elements within the 6G network. Moreover, we have analyzed the impact of feature extraction model partition points on the user experience of CVL within the proposed framework, and have introduced a RL-based joint optimization method for communication, computation, and confidentiality. In the case study, we have verified the proposed split-inference framework and joint optimization method. Finally, we have discussed potential directions for future research.


\bibliographystyle{IEEEtran}
\bibliography{bibRef}

\end{document}